# THE USE OF E - PORTFOLIOS IN TEACHING AND ASSESSMENT

Chionidou – Moskofoglou Maria, Doukakis Spyros, Lappa Amalia

Dept. of Primary Education, University of the Aegean

*In this paper, we will initially go through the results of assessment in mathematics according to the international assessment programs PISA, TIMSS (2003), with respect to students' portfolios. Furthermore, we will present the forms and the ways of assessment and will focus on that assessment which refers to the use of e-portfolios.*

## INTRODUCTION

Students' assessment constitutes an important part of the educational process. Moreover, it is considered that it can contribute to teaching as well as to the measurement of its results (Black P. and William D., 2003).

Today, the large amount of information and the use of electronic means lead, legitimately, to the development of new teaching and learning strategies and, consequently, to corresponding systems of assessment and evaluation (Birenbaum, 1996, Chionidou – Moskofoglou, 2000) with several objectives. Birenbaum cites that students should develop specific competencies in mathematics, including:

- cognitive competencies, such as problem solving, critical thinking, formulating questions, searching for relevant information, making informed judgements, efficient use of information, conducting observations, investigations, inventing and creating new things, analysing data, presenting data communicatively, oral and written expression;
- meta-cognitive competencies, such as self-reflection and self-evaluation;
- social competencies, such as leading discussions and conversations, persuading, co-operating, working in groups, etc. and
- affective dispositions, such as for instance perseverance, internal motivation, responsibility, self-efficacy, independence, flexibility, or coping with frustrating situations (Birenbaum 1996, p. 4).

## ACHIVEMENT RESULTS OF PISA AND TIMSS 2003

According to the study of PISA (2003), it becomes evident that the logic of assessment and the means that are used vary, to a great extent, from country to country. In our days, to a greater extent than in the past, in a lot of countries, the policy of assessment is orientated by the government or by a relevant public institution, aiming at the existence of concrete criteria so that qualitative work is achieved in the school units. The approaches and the ways of assessment are differentiated, depending on the definition of sought objectives and the formulation of the expected results in well-defined teaching units. PISA research data (2003) shows that the methods of assessment that are used in the countries participating in

OECD include standardised tests; tests developed by the teacher; teachers' judgmental ratings; students' portfolios; students' assignments, projects and homework.

As derived from PISA research data (2003), the students' portfolios are used to assess the performance of 43% of the students at least three times per year. Extending the aforementioned data, TIMSS (2003) suggests that the teachers of the participating countries have various attitudes towards the students' homework. On one hand, there are those teachers who assign their students homework of a relatively long duration (it exceeds 30 min.) and these assignments are frequent. On the other hand, there are those teachers who assign their students homework of a short duration (it does not exceed 30 min.) and these assignments are not frequent. An important data of the research concerns the way teachers face and develop their students' work; as concluded from the results of the data, teachers do not appear to have converging practices. Based, therefore, on the data:

- 78% of the teachers report that they check if their students have completed their work.
- 50% of the teachers collect, check, correct and return their students' worksheets and provide them with feedback of the appropriate solutions and/or make observations.
- One third of teachers declare that their students always-or almost always- correct their work in the classroom.

Finally, regarding the practice of assessing homework, one quarter of students on average receive scores on their completed work.

However, as indicated in PISA results (2003), the students' portfolios do not reveal any relation between the performance in PISA assessment instruments and their performance at school. Thus, it should be pointed that these forms of assessment require further research and analysis.

In Greece, relying on PISA data, the results are differentiated in several dimensions. The following table cites two ways of assessment; the rates that refer to the use of various methods in Greece; and the mean in the countries of OECD, as well as the corresponding students' scores:

Students' Portfolios

|        | 2 times per year or less | | At least three times per year | |
|--------|------|-----|------|-----|
| Greece | 83,0 | 449 | 17,0 | 431 |
| Mean   | 56,7 | 503 | 43,3 | 495 |

Student Homework and/or Projects

|        | | | | |
|--------|------|-----|------|-----|
| Greece | 85,3 | 447 | 14,7 | 435 |
| Mean   | 14,1 | 477 | 85,9 | 503 |

*Table 1. Methods of assessment and students' scores in mathematics (PISA, 2003)*

Taking into consideration the aforementioned data, Greece is below the mean of classification and, in specific, is placed 32nd among the 40 countries that participated. The data indicate that in Greece the student portfolios and work taking place at school and/or homework to assess 15-year-olds students are used rarely or not at all, as approximately 85% of teachers of mathematics.

During the last decade, the student's assessment was of primary importance to the Greek governments, too. Changes have been made through presidential decrees and ministerial decisions in order for the students to penetrate into the way of their learning, to develop strategies with a view to learn how to learn and, simultaneously, to develop lifelong competencies of learning (Chionidou-Moskofoglou M., Doukakis S., Mantzanaris K., 2005).

Although, however, the objective has been placed, teachers do not have the appropriate assistance in order to attain the expected goals. Therefore, whereas the directives exist and certain aims are achieved, what happens during the process of teaching is missing. According to Black P. and William D. (1998), these successful or unsuccessful objectives depend on the effectiveness of teaching.

In the next paragraphs we will try to show that the diversity of assessment methods sweeps along teaching and learning towards improvement and we will propose the use of portfolios as the means which can include those elements that bring out the quality of students' learning.

## ASSESSMENT AIMING AT LEARNING

## FORMS OF ASSESSMENT

The new Greek Cross-thematic Curriculum for Compulsory Education (Pedagogical Institute, 2003) reports that: "Assessment includes three important levels, with each one achieving certain objectives. The initial/diagnostic assessment is primarily used at the beginning of the learning process- and sometimes also during it- to identify the level of pupils' knowledge and interests and to identify possible difficulties that they face. The formative assessment is applied during teaching and is mainly of an informative character as it is used to monitor pupils´ progress towards the achievement of specific educational targets and the final/summative assessment is used to summarise and offer feedback, in order to evaluate to what extent the teaching and pedagogical targets have been achieved, in relation to the predefined goals".

## THEORIES OF LEARNING AND ASSESSMENT

According to Black P. and William D. (1998), learning is driven by what teachers and pupils do in classrooms, where teachers have to manage complicated and demanding situations, challenging the personal, emotional and social pressures amongst a group of 30 or so youngsters in order to help them to develop strategies with a view to learn how to learn, as well as develop lifelong competencies of learning (Chionidou-Moskofoglou M., et al., 2005). Furthermore, Black P. and William D. (1998) support that formative assessment is the heart of effective teaching and they provide data, which indicate that the improvement of formative evaluation can lead to better learning results.

The two basic learning theories can coexist in formative assessment. Behaviorist and Constructivist learning theories can be practically used for the purposes of a successful learning process. Moreover, the theory of Multiple Intelligences, can contribute to the ways that students learn. The application of the Behaviorist theory in assessment aims at discovering if the student has learned or if he/she can accomplish a predetermined task. Nevertheless, the application of the Constructivist theory in assessment can reveal what the student knows, what he/she comprehends and what he/she can do. (Forrester D. and Jantzie N., 2002).

**TECHNIQUES OF ASSESSMENT AND THEIR USE**

Students' assessment is an integral part of the educational process. It helps determine to what extent the teaching objectives, which are specified by the existing curriculums of the corresponding courses, have been achieved.

Assessment should combine various techniques in order for a valid, reliable, objective and unimpeachable evaluation of knowledge, students' critical thinking and dexterities to be achieved; and with a view to contribute to students' self-knowledge and to their objective information, depending on their individual learning level and potentials (Chionidou–Moskofoglou M. and Doukakis S., 2004).

Moreover, it should provide the teacher with the results of his/her work and provide teaching practice with feedback, aiming at the continuous improvement and increase of its effectiveness. To sum up, it should also contribute, to a great extent, to the informing of students' parents and guardians on their progress.

A vital necessity, according to Black P. and William D. (1998), concerns the students' active involvement in educational activities, with a view to collaborate and support the process of their own education. For the purposes of the aforementioned requirements, we propose the use of individual, for each student, portfolios during his/her mathematics teaching practice.

**PORTFOLIOS**

There are many types of portfolios, prepared for different purposes and using various resources. With regard to the use and application of portfolios to the educational environment, portfolios are structured to cover different needs, as these results from the educational process. Thus, it is supported that portfolios (Mason R., Pegler A. and Weller M., 2004) are those that:

- Offer a rich and textured view of a student's learning and development,
- Represent the outcomes of the student's overall progress and practice, with a view to create a personal CV. The innate assessment of the competence obtained in the course of work and learning (adaptability, communication, reasoning, problem solving, self-esteem and autonomy) is an important dimension of the student's image and can be evaluated through portfolios.
- Create a body of work that represents student's learning over the course of his/her education,
- Provide a valid result of the student's assessment.

Finally, portfolios constitute an accessible information repository, which can ensure the student's as well as the teacher's assessment and progress/attitude. Student's and teacher's self-assessment (Valli L. and Rennert–Ariev P., 2002) provides the possibility of improvement of the educational practice.

**CREATING PORTFOLIOS**

Students' portfolios should be applied in areas that are difficult to be taught and assessed through compatible methods. Such regions are, for instance: measurements, graphic representations, drawing of diagrams. In addition, portfolios should be created at the beginning of the school year, to provide an overall picture of the students´ competence and progress. Each student is indebted to work on subjects that he/she prefers in order to be able to monitor and reason his/her own work. Furthermore, students are challenged by tasks assigned by their teachers, who verify if these tasks are accomplished according to their own criteria (Woodward H., 2000). Portfolios follow a developmental process: collection, selection, reflection, projection and presentation of their data (Barret H., 2004).

**GOOD PORTFOLIO PRACTICES**

As presented in the figure, portfolios may include a varied material.

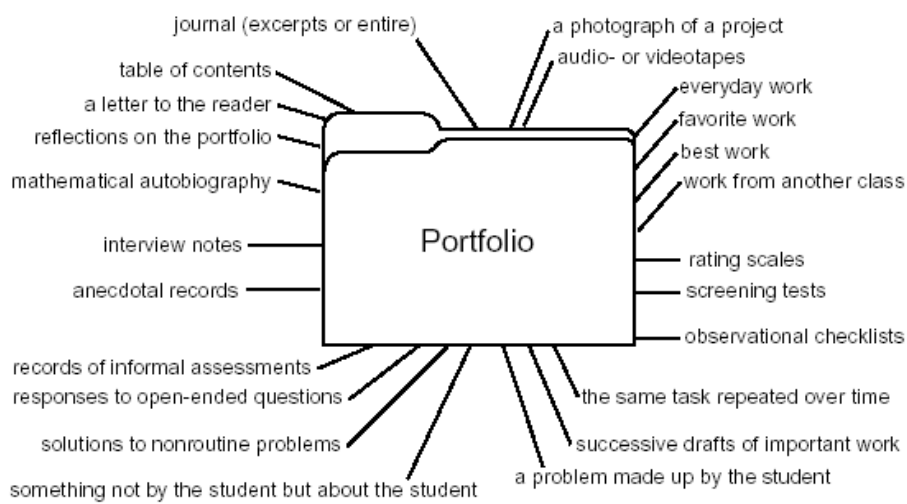

*Figure 1. Possible items in a mathematical portfolio (Kuhs T., 1994).*

However, they should include such activities and assignments that would help students to learn (Vosniadou P., 2001) and that:

- Allow and encourage students' active and constructive involvement.
- Provide students with social activities and enlarge students' participation in social school-life through collaborative activities.
- Establish links to real life and are relevant to the student's culture.
- Activate the acquisition of new knowledge and use the student's prior knowledge as a starting point.
- Lead to the employment of effective and flexible strategies.
- Assist students in creating and monitoring their own learning; setting their own goals; identifying false beliefs and misconceptions; correcting errors.

- Facilitate students to solve their internal inconsistencies and reconstruct their prior knowledge and beliefs, where necessary.
- Aim towards understanding rather than memorization.
- Challenge students to reflect and apply their knowledge.
- Provide considerable time and periods of practice and they are not pressing.
- Create motivated learners.

**E-PORTFOLIOS AND ELECTRONIC ASSESSMENT**

Although the use of Information and Communication Technology (ICT) is considered as a modern way of approaching teaching, learning and assessment, it has not been widely developed yet. In our days, the use of technologies in mathematics is distributed to the use of:

- Graphing Calculators,
- Computer Algebra Systems and educational software, namely: The Geometers' Sketchpad, Fathom Dynamic Statistics, Graph, E-Slate and so forth,
- text-processor, spreadsheet, presentations software, Internet,
- assessment through multiple-choice items, matching and True-False items.

These modern forms of technology are incorporated in e–portfolios, having two positive consequences: students are introduced to modern technologies and ICT is utilised in teaching practice (Chionidou-Moskofoglou et al. 2005).

In addition, it is incumbent upon ICT and, consequently, upon e –portfolios to support teachers who would attempt to attain particular educational objectives, but come up against difficulties in achieving them through compatible teaching methods. NCTM (2000) supports that: "through the use of technological tools, students might justify more general subjects, modelise and solve complicated problems, which could not be solved in the past".

E-portfolios are separated into two basic categories. In the first category, there are the technological tools of general use, text-processor, web-pages, multimedia tools and so forth. In the second category, there are the adapted systems that include servers, programming and databases.

E-portfolios allow students and teachers to collect and organise their files with many and multiple means (acoustical, visual, graphical, text). Web pages and contacts are used to organise the material and the data are connected on the bases of predetermined models.

One of the reasons that lead to the use of e-portfolios is that they are easily accessed; through them, the results of educational activities can be easily supervised. Moreover, copies are easily produced; there is portability; they occupy limited disk space; and the student is at the centre of the educational process. In the case of portfolios using databases through internet connection, the teacher, the guardian and the student can obtain a direct access and send comments on some work or on the content of the file electronically. Per regular time intervals, they can commend on the quality of the file and on its best and worst points.

For e-portfolios to be developed, a new culture is required as far as it concerns so much teachers' and students' education what for teachers' and students' support; further research into the precise planning of e-portfolios, a planning which should correspond with the one that supports "paper and pencil" portfolios; a training specialised in the use of information, communication and technology; the appropriate methodology in order for the student to respond to what he is asked and not only to what he/she can give; and teachers' further training. To sum up, issues of software and content security, connectivity, grouping, deployment and portability should be clarified before any use of e – portfolios.

**OUR FURTHER RESEARCH**

It is well known in mathematics research community, that, according to PISA 2003 report (p. 231): *"…for the use of student portfolios the relationship tends to be weaker and no clear pattern emerges. It is difficult to shed light on this with the limited information that PISA provides, and an examination how the use of different assessment instruments can contribute to raising performance levels requires further research and analysis"*. Moreover Beevers E. and Patersonthe S. (2002) side that *"…in the e–assessment of problem solving in mathematics, five theories are particularly relevant. The theories and their founders are shown below.*

- *Problem Solving Strategies: Polya*
- *Mathematical Problem Solving: Schoenfeld*
- *Information Processing: Miller*
- *Conditions of Learning: Gagne*
- *Operant Conditioning: Skinner"*.

We also added the Structure of Observed Learning Outcomes (SOLO) taxonomy: Biggs and Collis (1982).

Taking into consideration the aforementioned data, our research purpose is to study what kind of assessment techniques, according to assessment theories, should be included in an e-pupil portfolio so as to be used in an optimised way. As a result, we put the following research questions:

- In which way can all the various techniques (qualitative and quantitative) be used in the e-portfolio?
- How will the e-portfolio, using all various (qualitative and quantitative) techniques, give a valid, reliable, and unimpeachable assessment of students' knowledge?
- What is the relation between the students' learning and knowing mathematics and e-portfolios?

E-portfolios appear to be a useful research challenge. They should constitute a supportive, dynamic, formative and pleasant assessment environment, which has the potential for following them throughout their life giving meaning to their mathematical knowledge.